\documentclass[citeautoscript,twocolumn,showpacs,superscriptaddress,prb,aps,floatfix]{revtex4}
\usepackage{graphicx}
\usepackage{rotating}
\usepackage{amsmath}
\usepackage{url}
\usepackage{bm}
\usepackage[dvips]{color}
\usepackage{subfigure}
\newcommand{\bk}{{\bm k}}

\newcommand{\bq}{{\bm q}}
\newcommand{\EF}{$E_F$}

\newcommand{\nk}{{n\bm k}}
\newcommand{\nkp}{{n'\bm k'}}

\newcommand{\tc}{$T_c~$}
\newcommand{\Dk}{\Delta_{n{\bm k}}}

\newcommand{\gkkp}{|g^{nn'}_{{\bm k,\bm k'},\nu}|^2}

\newcommand{\be}{\begin{equation}}
\newcommand{\ee}{\end{equation}}
\newcommand{\cabesi}{CaBeSi}
\newcommand{\cabesix}{CaBe$_x$Si$_{2-x}~$}
\newcommand{\mgbtwo}{MgB$_2$}
\newcommand{\albtwo}{AlB$_2$}
\newcommand{\s}{$\sigma$}
\newcommand{\p}{$\pi$}
\newcommand{\pa}{$\pi_a$}
\newcommand{\pb}{$\pi_b$}

\newcommand{\us}{\sigma_1}
\newcommand{\ds}{\sigma_2}
\newcommand{\ant}[2]{\;\;\begin{matrix}{\bf #1}\\(#2)\end{matrix}\;}
\begin{document}

\title{Electronic, dynamical and superconducting properties of CaBeSi}

\author{C.~Bersier} 
\affiliation{Institut f{\"u}r Theoretische Physik, Freie Universit{\"a}t
Berlin, Arnimallee 14, D-14195 Berlin, Germany}
\affiliation{European Theoretical Spectroscopy Facility (ETSF)}
\author{A.~Floris} 
\affiliation{Institut f{\"u}r Theoretische Physik, Freie Universit{\"a}t
Berlin, Arnimallee 14, D-14195 Berlin, Germany}
\affiliation{European Theoretical Spectroscopy Facility (ETSF)}
%\affiliation{INFM SLACS, Sardinian Laboratory for Computational Materials
%Science and Dipartimento di Scienze Fisiche, Universit\`a degli Studi di Cagliari,
%S.P. Monserrato-Sestu km 0.700, I--09124 Monserrato (Cagliari), Italy}

\author{A.~Sanna} 
\affiliation{Institut f{\"u}r Theoretische Physik, Freie Universit{\"a}t
Berlin, Arnimallee 14, D-14195 Berlin, Germany}
\affiliation{European Theoretical Spectroscopy Facility (ETSF)}
\affiliation{INFM SLACS, Sardinian Laboratory for Computational Materials
Science and Dipartimento di Scienze Fisiche, Universit\`a degli Studi di Cagliari,
S.P. Monserrato-Sestu km 0.700, I--09124 Monserrato (Cagliari), Italy}
\author{G.~Profeta}
\affiliation{CNISM - Dipartimento di Fisica,
Universit\`a degli Studi dell'Aquila, Via Vetoio 10,
I-67010 Coppito (L'Aquila) Italy}

\author{A.~Continenza}
\affiliation{CNISM - Dipartimento di Fisica,
Universit\`a degli Studi dell'Aquila, Via Vetoio 10,
I-67010 Coppito (L'Aquila) Italy}

\author{E.\,K.\,U.~Gross}
\affiliation{Institut f{\"u}r Theoretische Physik, Freie Universit{\"a}t
Berlin, Arnimallee 14, D-14195 Berlin, Germany}
\affiliation{European Theoretical Spectroscopy Facility (ETSF)}
\author{S.~Massidda} 
%\altaffiliation{Also at INFM-LAMIA, Genova, Italy}
\affiliation{INFM SLACS, Sardinian Laboratory for Computational Materials
Science and Dipartimento di Scienze Fisiche, Universit\`a degli Studi di Cagliari,
S.P. Monserrato-Sestu km 0.700, I--09124 Monserrato (Cagliari), Italy}

\date{\today}

\begin{abstract}
We report first-principles calculations on the normal and 
superconducting state of \cabesix\ ($x=1$),   in the framework of 
density functional theory for superconductors (SCDFT). \cabesi\ is isostructural and 
isoelectronic to \mgbtwo\ and this makes possible a direct comparison 
of the electronic and vibrational properties and the electron-phonon 
interaction of the two materials. Despite the many similarities with \mgbtwo\
({\em e.g.} \s\ bands at the Fermi level and a larger Fermi surface nesting), 
according to our calculations  
 \cabesi\ has  a very low  critical temperature (\tc$\approx 0.4$~K, consistent with the experiment).
 \cabesi\  exhibits a complex gap structure, with three gaps at Fermi level: 
besides the two \s\ and \p\ gaps, present also in \mgbtwo,  the 
appearance of a third gap is related to the anisotropy of the 
Coulomb repulsion, acting in different way
on the bonding and antibonding electronic \p\ states.

\end{abstract}
\pacs{74.25.Jb,74.25.Kc,74.70.Ad}
\maketitle

\section{Introduction}

The complexity and the fragility of the effective pairing interaction 
-- result of a fine interplay of opposite contributions -- 
and its non--trivial dependence on chemical properties
make the search of new superconducting materials  a very difficult task,
even within the class of  phonon mediated superconductors.
%The search of new superconducting materials is a difficult task, 
%even within the class of  phonon mediated superconductors,
%both because of the weakness of the effective interaction 
%-- result of a fine interplay of opposite contributions -- 
%and of its non--trivial dependence on chemical properties.  
As a result, research often goes along the lines of searching in the 
``neighborhood'' of known superconductors. This has been the case 
of \mgbtwo\ that owes its superconducting (SC) properties\cite{an-pickett,liu,kong} 
essentially to the presence of holes in the $sp^2$ $\sigma$ B-B covalent 
bonds, strongly coupled with the $E_{2g}$ stretching mode.
During the last years, \mgbtwo\ boosted experimental and theoretical  research
in the class of diborides, layered, graphite-like materials, 
all sharing some hopefully relevant features with the ``parent
compound''. 
Although this effort did not succeed in finding  
superconductors with better  (or at least equivalent) properties than
\mgbtwo, it brought to the experimental discovery of some  new superconductors 
like graphite intercalated compounds\cite{gic1,gic2}, boron doped diamond\cite{diamond}, and to  some theoretical proposals\cite{LIBC,BC3}.

%However, in spite of a huge theoretical and experimental effort, 
%up to now no phonon mediated materials have been found having 
%SC properties comparable to the ones of \mgbtwo\, 
%although some materials have been proposed\cite{LIBC,BC3}.
Among \mgbtwo-like materials, one interesting case is 
represented by CaSi$_2$, which at ambient pressure has a 
rhombohedral structure and changes to a trigonal phase between 
$\approx$ 7 and 9.5 GPa. 
At $P > 16$~GPa, CaSi$_2$ has the {\it \albtwo} structure and SC 
critical temperatures ({$T_c$}) up to 14~K. 
The stability of the {\it \albtwo} phase at ambient pressure is 
achieved through hole doping.
 The stabilization process has been related\cite{satta}  to the filling of 
the \p\ antibonding bands (\pa). 
 Let us consider a material 
CaX$_x$Si$_{2-x}~$  (X having a lower valence than Si): 
increasing $x$ from zero introduces holes in the
\pa\ bands, with a change in the X-Si bond that from an
$sp^3$-like character acquires a progressive $sp^2$ character, 
stabilizing the {\it \albtwo} structure\cite{satta}.
% with a consequent distortion known in the literature as 'trigonal distortion'. 
Thus, hole-doped CaSi$_2$ could represent (with some differences) a good 
candidate for an
\mgbtwo-like material, although its stabilization at high pressure poses some
problems.
Nevertheless, experimentally, the {\it \albtwo} phase was observed at 
ambient pressure by doping CaSi$_2$ with Al\cite{imai,lorenz} or, 
in a much less studied case, with Be \cite{may,sano}. 
Recently, \cabesix was investigated in the doping range  
$0.5 \leq x \leq 1$ \cite{sano} and was observed to 
have the {\it \albtwo} structure at $x=0.75$ \cite{may,sano} 
and, against the expectations, not to be a superconductor down to  
4.2 K \cite{sano}. 
%For $x\neq0.75$ other sub-phases, different from  {\it \albtwo}, 
%were also  observed \cite{may,sano}. 

The goal of this paper is to clarify why this happens 
%ANTO try to understand why this happens 
despite the presence of \s\ holes, as in \mgbtwo, and an even more
pronounced nesting. 
We investigate  the normal and 
 SC phase of \cabesix at $x=1$ 
 (\cabesi\ in the following)\cite{nota}, in the {\it \albtwo} phase, 
 with Be and Si atoms alternating within the honeycomb layers with 
 an $AA$ stacking. 
%The chosen doping $x=1$  is not far from the experimental 
%stoichiometry ($x=0.75$) and  avoids  cumbersome
%supercell approaches, and problems related to 
%Be and Si ordering,  not affecting the conclusions of the work.

The paper is organized as follows: 
In Secs. \ref{scdft} and \ref{comp} we sketch our computational framework and
 describe the computational details of our calculations. 
 In Sec. \ref{normres} we illustrate the normal state properties 
 and the e-ph coupling of \cabesi\ .  
 The superconducting properties are discussed in Sec. \ref{superres}. 
% although we predict a low \tc, the SCDFT {\bf k}-resolved effective 
% interaction has a complex structure, being more anisotropic than in \mgbtwo. 
% This brings to a three-gap superconductivity, 
% that is interpreted through the use of a toy model.  
 Finally, in Sec. \ref{concl}, we  summarize our results.

\section{DENSITY FUNCTIONAL THEORY FOR SUPERCONDUCTORS (SCDFT)}
%\subsection{Density Functional Theory for Superconductors (SCDFT)}
\label{scdft}

The superconducting properties are studied within the density functional 
theory for superconductivity (SCDFT) \cite{noiI,noiII}, 
a completely parameter free approach which allows 
to predict the SC gap and \tc values of real materials \cite{noiI,noiII}.
As the SCDFT method was discussed in full detail in previous 
papers\cite{noiI,noiII}, 
here we only sketch the main points concerning the theory, 
whose central result  
is the generalized, self-consistent, gap equation:
\begin{equation} 
  \label{gapeq} 
  \Delta_\nk = - {\cal Z}_\nk \;\Delta_\nk 
  -\frac{1}{2}\sum_\nkp {\cal K}_{\nk,\nkp}
  \frac{\tanh\left(\frac{\beta}{2}E_\nkp\right)}{E_\nkp} \; \Delta_\nkp,
\end{equation}
where $n$ and $\bk$ are respectively the electronic band index and the Bloch wave vector; $\Dk$~ is the SC gap function;  $\beta$ is the inverse temperature; $E_\nk=\sqrt{(\varepsilon_\nk-\mu)^2+ \left|\Delta_\nk\right|^2}$ are the SC quasiparticle energies, defined in terms of the gap 
function, the Kohn-Sham eigenenergies $\varepsilon_\nk$ of the normal state and the chemical potential $\mu$. The {\it universal} kernel ${\cal K}_{\nk,\nkp}$~ 
appearing in Eq.~(\ref{gapeq}) consists of two contributions~ 
${\cal K}={\cal K}^{\rm e-ph}+{\cal K}^{\rm e-e}$, representing the effects of 
the e-ph and the e-e interactions respectively. ${\cal K}^{\rm e-ph}$~ 
is temperature dependent and involves the e-ph coupling matrix elements $\gkkp$ and the phonon spectrum $\omega_{\bq\nu}$,
while ${\cal K}^{\rm e-e}$ contains the matrix elements 
of the screened Coulomb repulsion. 
%Eq.~(\ref{gapeq}) has the same structure 
%as the BCS gap equation, with the kernel ${\cal K}$ replacing the model interaction 
%of BCS theory. This similarity allows us to interpret ${\cal K}$ as an effective 
%interaction, responsible for the binding of the Cooper pairs. 
Finally, the {\it universal} diagonal 
(and temperature dependent) term ${\cal Z}_\nk$ plays a similar role as the renormalization term in the Eliashberg equations. We emphasize that Eq.~(\ref{gapeq}) is not a mean-field equation (as 
in BCS theory), since it contains correlation effects via the SC exchange-correlation (xc) functional entering ${\cal K}$ and ${\cal Z}$. Furthermore, it has the 
form of a static equation -- i.e., it does not depend {\it explicitly} on the frequency --  
and therefore has a simpler structure (and computationally more manageable) 
than the Eliashberg equations. However, this 
certainly does not imply that retardation effects are absent from the theory. Once again, retardation effects
enter through the xc functional, as explained in Refs. ~\onlinecite{noiI,noiII} .

An important feature of the SCDFT approach is the capability to include  the $\nk$-resolved Coulomb repulsion ${V}^{\rm e-e}$
{\it ab-initio}, without any adjustable parameter. The Coulomb interaction was screened 
 with a static dielectric matrix, within the Random Phase Approximation (RPA)\cite{self}(see below).  

%As we will see, in \cabesi\ the inclusion of $\nk$-resolved interaction versus an average one, allows
%the appearance of  a much more complex  structure of the SC gap.

\section{COMPUTATIONAL DETAILS} 
 
\label{comp}

The electronic band structure $\varepsilon_{\nk}$, 
the phonon spectrum $\omega_{\bq\nu}$, and the electron-phonon (e-ph)
 and Coulomb 
matrix elements (ME) with respect to the Bloch functions necessary 
to solve  the SC gap  
(Eq. \ref{gapeq}), were obtained within the 
planewave-pseudopotential method \cite{pwscf}. 
We used Troullier-Martins\cite{tm}, 
norm-conserving pseudopotentials\cite{fhi}, 
with 2s, 3s-3p and 3s-3p-4s for Be, Si and Ca, respectively,
in the GGA-PBE\cite{pbe} approximation for the xc functional. 
The electronic self-consistent cycle was performed with a 60 Ry energy 
cut-off, a $18^3$ Monkhorst-Pack $\bk$-point mesh. 
%and a smearing parameter of 0.01 Ry.

The optimized lattice parameters (at $x=1$ Be doping) are 
$a=3.895~$\AA~ and $c/a= 1.112~$, whereas the experimental values (at $x=0.75$) are $a=3.94$~\AA~ and $c/a= 1.112$ \cite{may,sano}. 
Our obtained, slightly smaller GGA-PBE $a$ constant is 
justified by the larger amount of Be (with smaller covalent radius than Si) 
present in the calculated system. 
%(at the same doping GGA's functionals would tend to slightly {\it overestimate} the lattice parameters). 
%The pseudopotential band structure is in good agreement with previous calculations \cite{satta} and with a test  calculation performed within the all-electron FLAPW method \cite{exciting}.

 Phonons and e-ph ME were calculated via 
density functional perturbation theory \cite{baroni}. 
The phonons were computed on the irreducible set of a regular mesh 
of $8^3$  $\bq$-points, and a  $16^3$ Monkhorst-Pack $\bk$-points for
electronic integration, with a smearing parameter of 0.35 eV. 
These parameters were sufficient to achieve  convergence within 0.5~meV in the 
frequency of the 
$E_{2g}(\Gamma)$ mode. 
Our calculated  $\omega_{E_{2g}}(\Gamma)=59.0~$meV compares well with the value
$\omega_{E_{2g}}(\Gamma)=57.1~$meV obtained by frozen phonon calculations
in Ref. ~\onlinecite{satta}, 
computed at the slightly larger $a=3.914$~\AA.  
The e-ph ME were calculated on the same $\bq$-points grid as the phonons, 
and on a denser grid of  $24^3$ $\bk$-points, while the RPA screened 
Coulomb ME were calculated\cite{self} on a mesh of 
$9^3\;\times\;9^3$ $\bk$ and ${\bf k'}$-points.
The SC gap function is extremely peaked around the Fermi
surface (within the characteristic phonon energy), whereas at 
higher energies it is rather smooth 
(and it changes sign, due to the e-e interaction). 
This implies that a converged solution of Eq. (\ref{gapeq}) 
needs a denser $\bk$-points sampling around \EF, while a coarser one is
sufficient elsewhere. This highly non uniform mesh of the Brillouin zone 
is realized with $8\times10^3$   and $500$ independent $\bk$-points for 
bands crossing and not crossing the Fermi level respectively. 
Finally, 15-20 self consistent iterations were sufficient 
to achieve a complete convergence of the gap.

%\begin{table*}
%\begin{center}
%{\large
%\begin{tabular}{ccccccc}
%
%\hline
%\hline
%  &  \hspace{0.5cm} $a$ & \hspace{0.5cm}$c$  & \hspace{0.5cm}  Si-Si & \hspace{0.5cm} Si-Be  &\hspace{0.5cm}  B-B & \hspace{0.5cm} $\frac{A_A}{A_{\Gamma}}$ \\ 
%\hline 
%\cabesi  & \hspace{0.5cm} 3.895 & \hspace{0.5cm} 4.331 & \hspace{0.5cm} 3.895   &\hspace{0.5cm} 2.249    &  \hspace{0.5cm}-  &\hspace{0.5cm} 5.4, 2.4  \\  
%\mgbtwo\footnote{Experimental constants}  & \hspace{0.5cm} 3.083 & \hspace{0.5cm} 3.52 & \hspace{0.5cm} -   &\hspace{0.5cm} -    &  \hspace{0.5cm}1.780  &\hspace{0.5cm} 2.21, 2.07  \\  
%\hline 
%\end{tabular}
%}
%\caption{Structural parameters (in \AA) of \cabesi\ and \mgbtwo. The two values on the last column refer
%to internal ($\us$) and external ($\ds$ ) \s\ bands respectively.  \label{tab1}}
%\end{center}
%\end{table*}

\begin{table}
%\begin{center}
%{\large
\begin{tabular}{cccccc}

\hline
\hline
  &  \hspace{0.3cm} $a$ & \hspace{0.3cm}$c$  & \hspace{0.3cm}  Si-Si & \hspace{0.3cm} Si-Be  &\hspace{0.3cm}  B-B \\ 
\hline 
\cabesi  & \hspace{0.3cm} 3.895 & \hspace{0.3cm} 4.331 & \hspace{0.3cm} 3.895   &\hspace{0.3cm} 2.249    &  \hspace{0.3cm}-  \\  
\mgbtwo\footnote{Experimental constants}  & \hspace{0.3cm} 3.083 & \hspace{0.3cm} 3.52 & \hspace{0.3cm} -   &\hspace{0.3cm} -    &  \hspace{0.3cm}1.780   \\  
\hline 
\end{tabular}
%}
\caption{Structural parameters (in \AA) of \cabesi\ and \mgbtwo. \label{tab1} } % The two values on the last column refer to internal ($\us$) and external ($\ds$ ) \s\ bands respectively. 
%\end{center}
\end{table}

\section{NORMAL STATE PROPERTIES}
\label{normres}

%The \cabesi\ electronic and vibrational properties will be discussed in strict comparison with the ones of \mgbtwo. We will emphasize  the similarities and  the strong differences that bring the two systems to have  very different SC properties. 

\subsection{Electronic structure and Fermi surface}

\begin{figure}[t]
  \begin{center}
    \includegraphics[clip,width=0.43\textwidth]{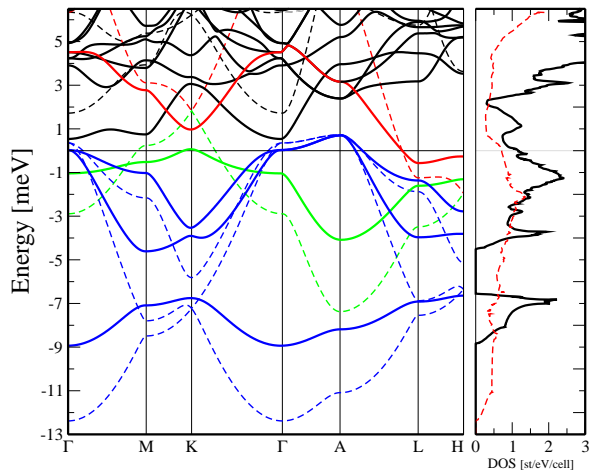}
%    \vspace{0.2cm}

%    \includegraphics[clip,width=0.43\textwidth]{fig2.eps}
  \end{center}

  \caption{(color online) Band structure and DOS of \cabesi\ (full lines) and \mgbtwo\ (dashed lines). Colors refer to different characters: \s\ (blue), \p\ bonding (green), \p\ antibonding (red).}
%{\bf CAN WE SEE IT IN BLACK AND WHITE?  WHICH STATES ARE Ca d? }}
  \label{bands}
\end{figure}

\begin{figure}[t]
  \begin{center}
    \includegraphics[clip,width=0.43\textwidth]{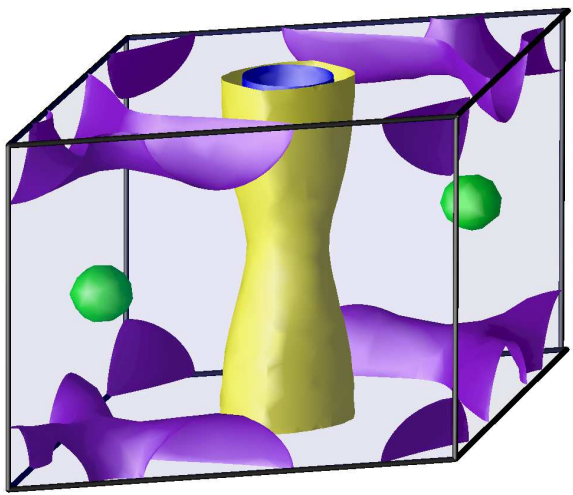}
%    \vspace{0.2cm}

%    \includegraphics[clip,width=0.43\textwidth]{fig2b.eps}
  \end{center}

%  \caption{(color online) \cabesi\ Fermi surface on the unit cell given by the reciprocal lattice vectors. Yellow and Blue tubular sheets are due to \s\ states. Violet sheet on the top and bottom of the cell are \pa\ states, while small green balls are due to the hole pocket in \pb\ states.}
  \caption{(color online) \cabesi\ Fermi surface. The colors identify the three gaps at \EF: the largest \s\ (yellow and blue cylindrical-like sheets); the intermediate \p\ bonding (green spheres); the lowest \p\ antibonding  (violet).}
  \label{fs}
\end{figure}

\begin{figure}[t]
  \begin{center}
\includegraphics[clip,width=0.23\textwidth]{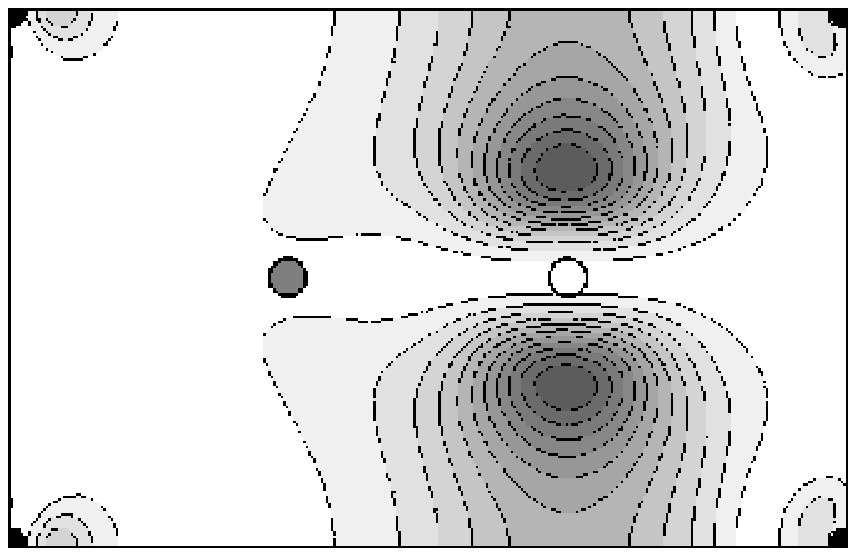}
\includegraphics[clip,width=0.23\textwidth]{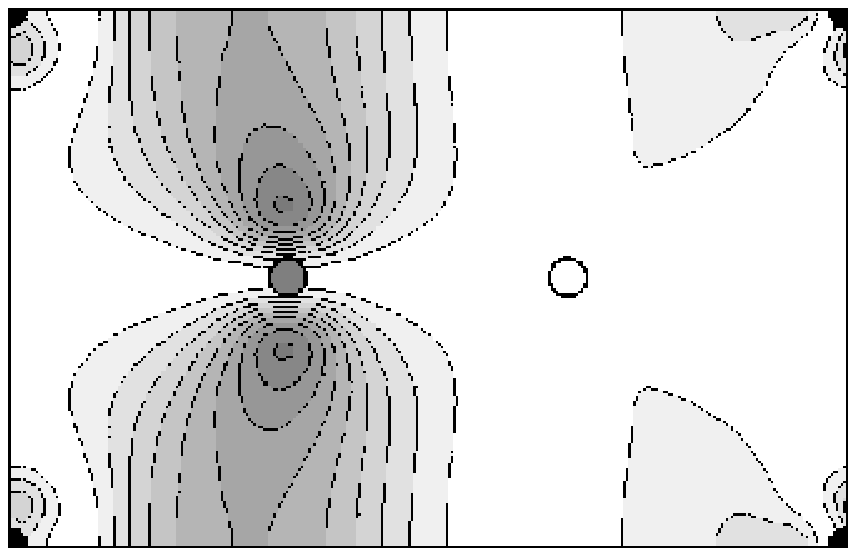}
\vspace{0.1cm}

\includegraphics[clip,width=0.23\textwidth]{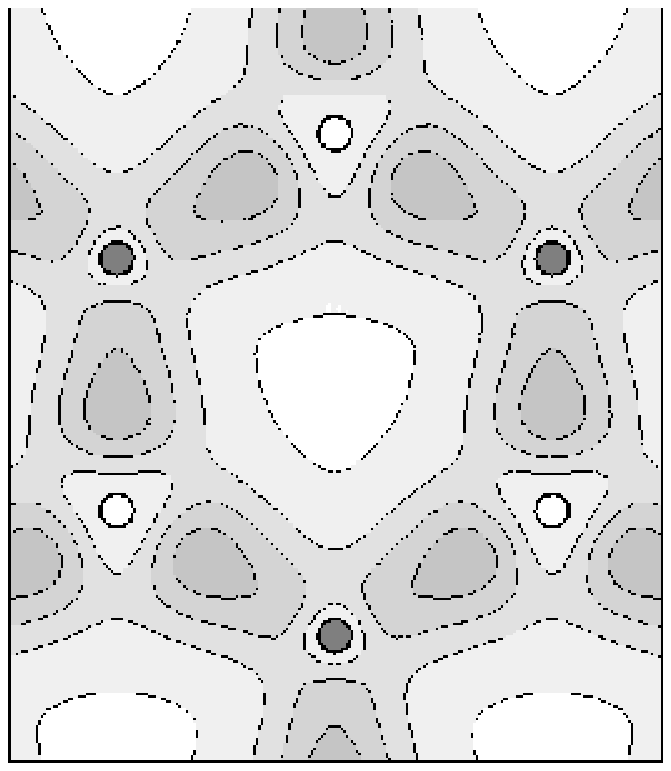}
\includegraphics[clip,width=0.23\textwidth]{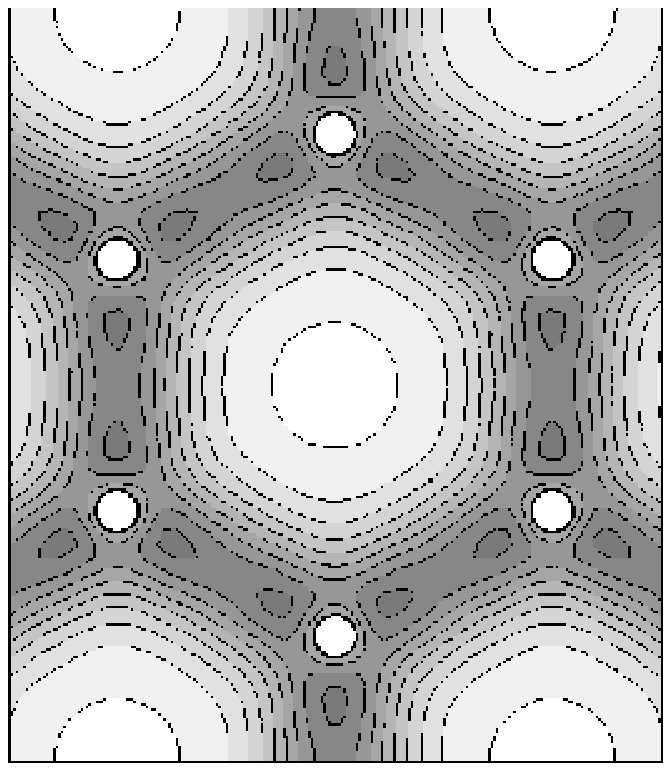}
\vspace{0.1cm}

\includegraphics[clip,width=0.23\textwidth]{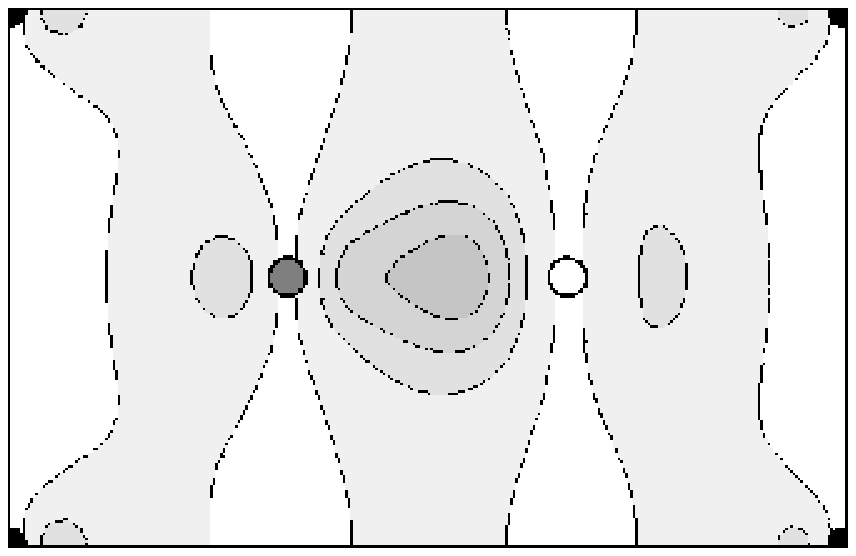}
\includegraphics[clip,width=0.23\textwidth]{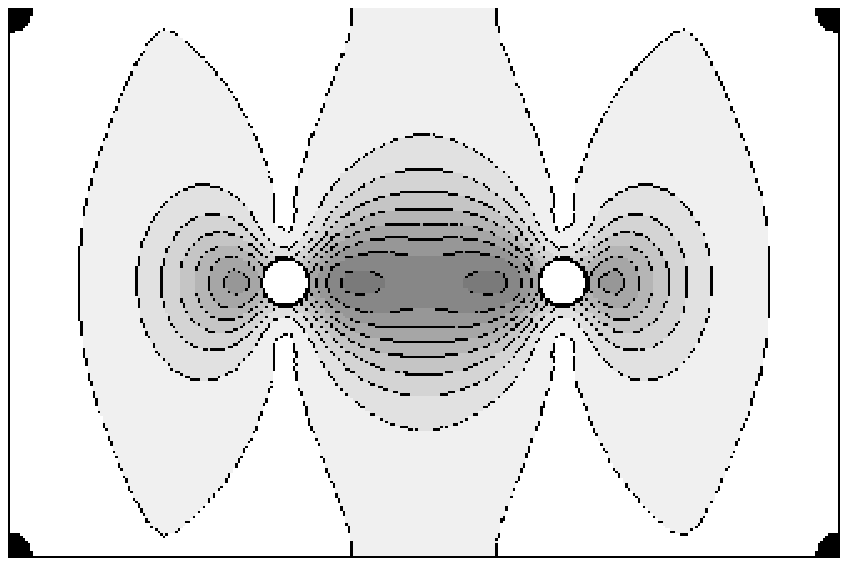}
%    \vspace{0.2cm}

%    \includegraphics[clip,width=0.43\textwidth]{fig2.eps}
  \end{center}

  \caption{\cabesi\ and \mgbtwo\ charge. 
  Upper panel: \cabesi\ \p\ bonding (left) and
\p\ antibonding (right) charge at the H point; empty 
(full) circle refers to Si (Be). Middle and lower panels:
 \cabesi\ (left) and \mgbtwo\ (right) \s\ charge at the $\Gamma$ point;
 in the middle (lower) panel, the z direction is perpendicular (parallel) to the page.}
  \label{charge}
\end{figure}

Fig.~\ref{bands} shows the \cabesi\ and  \mgbtwo\ band structure. 
A general similarity is found: in both materials the \s, $\pi$-bonding (\pb) 
and $\pi$-antibonding (\pa) bands cross the Fermi level (\EF). 
As pointed out in Ref. ~\onlinecite{satta}, and similarly to
CaAlSi\cite{giantomassi}, the reduced symmetry (space group P\=6m2)
%No. 187 {\textcolor{red} toglierei il numero del gruppo} 
with respect to \mgbtwo\ (P6/mmm), related to the partially ionic nature of %  (P6/mmm No. 191)
the B--Si bond, 
implies the splittings of the \s\ and  the \pb-\pa\ bands 
at the K and H points of the Brillouin zone (BZ). In
\cabesi\ the \pb\ bands are almost fully occupied, leaving only 
small holes pockets at K 
 which give rise to the little \p\  spheres of the Fermi 
surface (FS) (see Fig.~\ref{fs}). These
replace the \p\ tubular structure present in \mgbtwo. 
The \pa\ bands are only partially occupied, allowing the stabilization of the 
  Si-Be $sp^2$ network against a  $sp^3$-like distortion. 
In fact, the latter  takes place in the presence of a larger amount of Si, 
$i.e.$ with a larger filling 
of the \pa\ bands \cite{satta}. 
%The \s\-FS  (cylindrical-like) and \pa\ 
%sheets are similar to the ones of \mgbtwo, although the \cabesi\ \s\ sheets 
%(mainly the internal \s) have a larger ratio of the extremal 
%areas at A and $\Gamma$ points, due to the larger $\Gamma$A \s\ dispersion.
 
Due to the different electronegativity of Si and Be, 
we expect a change in the charge distribution related to their bond, 
in comparison with the B-B bond in \mgbtwo. 
In Fig.~\ref{charge} (upper panel) we plot the 
\pb\ and \pa\ charge at the BZ H point. 
We see that the \pb\ (\pa) charge
 is clearly associated to the Si (Be) atom.  Having in mind the different 
occupation of the \pb\ and \pa\ bands, we conclude that 
there is a charge disproportion in favor of Si.
%  We will see (Sec. \ref{superres}) how this fact has important consequences in the \p\ and \ps\ intraband Coulomb ME values, giving rise to a complex SC gap structure, with three gaps at \EF.

Another important chemical difference between \mgbtwo\ and \cabesi\ is 
the presence of Ca-$d$ states at $\sim 5$ eV  above \EF\ that strongly interact 
with \p\ and \s\ bands, in different ways along the zone. 
This contributes to the reduced in plane \p\ bandwidth 
($\approx$ 1.1 eV and $\approx$ 2.8 eV along $\Gamma$MK  and ALH) in 
\cabesi\ with respect to \mgbtwo\ ($\approx$ 5.5 eV). 
In fact, while  at the $\Gamma$ point there is no interaction between 
\p\ states and the cation $d$ orbitals, the interaction is possible 
in the high symmetry points $A$, $M$ and $K$, therefore 
suppressing the large dispersion of these states observed in \mgbtwo. 
%{\bf *** I am not sure of the following, the Si orbitals are larger} S.M.
A further reason for the reduced \p\ bandwidth in \cabesi\ is 
a lower chemical pressure effect due to its larger unit cell in 
comparison to \mgbtwo.

%From the viwpoint of the the e-ph coupling, the \s\ bond deserves particular 
%attention.  ! Here there is no direct connection with the e-ph coupling. Until an explanation is given.
Looking at the \s\ charge in the two materials 
(Fig.~\ref{charge}, middle and lower panels), we notice 
a stronger localization in \mgbtwo\ along the B-B direction, 
whereas in \cabesi\ the \s\ charge has a clear ionic component and 
is more delocalized both in the in-plane and out-of-plane directions.
This makes the Si-Be bond much weaker than the B-B one affecting the strength of the e-ph coupling.
%As \mgbtwo\ owes its properties to the stenght of its bonds\cite{an-pickett}  in advance that strongly reduce the e-ph coupling.

As far as the \s\ bandwidth is concerned, 
%This produces dramatic effects on the \s\ bandwidth: in fact, 
the larger Si-Be distance (see Table \ref{tab1}) and a lower Si-Be interaction
(compared to B-B) explain the reduced in plane  \s\ -band dispersion 
(for the lower \s\,  $\approx 5$ eV and $\approx 8$ eV 
in \cabesi\ and \mgbtwo, respectively). 
On the other hand, the larger out of plane 
dispersion of \s\ bands in \cabesi\ ( $\approx 1 $ eV)  
versus \mgbtwo\ ($\approx 0.5 $ eV) is only partially justified with 
the larger $z$-extension  of the \cabesi\ \s\ charge 
(compensated by the $\approx 25\%$ larger interlayer distance), and is
 mainly related to the \s\ -- Ca-$d_{x^2-y^2}$ interaction allowed 
at the $\Gamma$ point but not at A (see  Fig.~\ref{bands}).  
%{\bf HOW MUCH P AND HOW MUCH d?}
Difference in band dispersion gives rise to much of the warping of the corresponding 
Fermi surface.

%Fig. \ref{charge} (lower panel)  shows also that the Si-Be \s bond 
%has a ionic component,
%due to the stronger electronegativity of Si. 

% The delocalization can be related in turn to
%the nature of the Si-Be bond, less covalent that the B-B bond. 
%As we shall see, the low degree of delocalization has important consequences
%for the strong reduction of the  E$_{2g}$ e-ph ME in \cabesi, compared to \mgbtwo.  

\subsection{Phonons and electron-phonon coupling}%: A comparison with \mgbtwo}

\begin{figure}[t]
  \begin{center}
    \includegraphics[clip,width=0.43\textwidth]{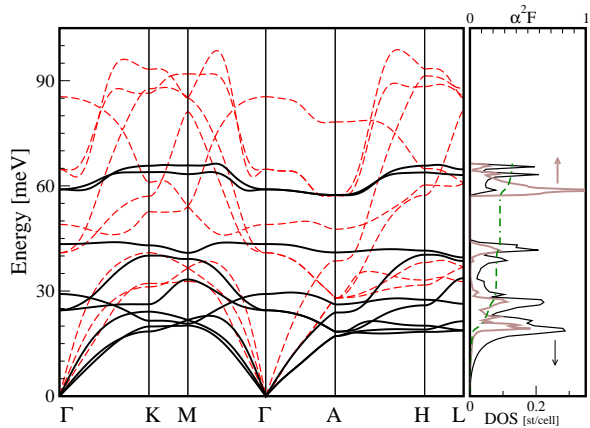}
%    \vspace{0.2cm}

%    \includegraphics[clip,width=0.43\textwidth]{fig2.eps}
  \end{center}

  \caption{(color online) Left panel: phonon structure of \cabesi\ (full lines) and \mgbtwo\ (dashed lines). Right panel: \cabesi\  phonon DOS (thin black line),  Eliashberg function (thick brown line) and integration curve $2\int_0^{\omega}\alpha^2F(\omega')/\omega' d\omega'$ (dot-dashed green line), whose final value is the total e-ph coupling $\lambda$.}
%and Eliashberg function}
  \label{phononeli}
\end{figure}

%\begin{figure}[t]
%  \begin{center}
%    \includegraphics[clip,width=0.43\textwidth]{fig1.eps}
%    \vspace{0.2cm}
%
%    \includegraphics[clip,width=0.43\textwidth]{fig2.eps}
%  \end{center}
% \caption{Nesting function and e-ph linewidths of \cabesi.}
%  \label{nestgamma}
%\end{figure}

The previous discussion shows that, despite the general similarities, \cabesi\ and \mgbtwo\
have rather different chemical and electronic properties. As expected, they
determine both the dynamical properties and the electron-phonon coupling.
In fact, \cabesi\  frequencies are lower (see Fig.~\ref{phononeli}) 
 in comparison with \mgbtwo, mainly due to the larger mass of Ca and Si versus 
 Mg and B. 

The E$_{2g}$ mode is fairly flat along the in plane BZ symmetry 
 lines and it shows only a very weak renormalization 
 along M$\Gamma$ and AH lines  (four times smaller than in \mgbtwo\ \cite{kong,bohnen,shukla}) 
due to the very small E$_{2g}$
electron-phonon matrix elements.
In turn, this is related to the delocalized and ionic nature
of the \s\ bonds in \cabesi\ (see Fig.~\ref{charge}). 
In fact, the connection between strongly covalent bonds  and 
strong e-ph coupling seems to be a general feature 
\cite{an-pickett,lial,noik,pb}.

The B$_{1g}$ mode lower than the  E$_{2g}$ everywhere in  the 
 BZ, without exhibiting the features found in CaAlSi, where it is
very soft, due to a strong  interband coupling between the interlayer and 
\pa\ states \cite{giantomassi}.

The strongly reduced e-ph renormalization of the \cabesi\ E$_{2g}$ mode is not related to poor FS nesting features in this material, but only to the small value of the e-ph ME themselves. In fact, we have calculated the \s\ nesting function \small {$\eta^\sigma_{ {\bm q} }= \frac{1}{N_\sigma(0)} \sum_{\bm k\in \sigma } \delta(\epsilon_{ \bm k })~ \delta(\epsilon_{\bm k+ \bm q})$} (where ${\bm k}+{\bm q}\in\sigma$ and $N_\sigma(0)$ is the \s\ DOS at \EF), obtaining {\it larger} values for \cabesi\ than for \mgbtwo\ (roughly a factor of 2).  
%This function enters directly in the phonon linewidths expressions\cite{allen72} 

%(the interlayer band crosses \EF in CaAlSi).
% and it  is responsible for most of the e-ph coupling in that material  \cite{giantomassi}. 

%This strongly reduced e-ph renormalization of the \cabesi\ E$_{2g}$ mode is not related to poor FS nesting features in this material: Fig.~\ref{nestgamma}a shows that the \cabesi\ in plane \s\ nesting function (renormalized with the \s\ density of states at \EF\ $N_\sigma$), has larger values than in \mgbtwo.  In 
%Fig.~\ref{nestgamma}b, we plot the E$_{2g}$ phonon linewidths\cite{allen72} of the two materials ATTNEANCHEMGB2?. Although the DOS at \EF\  is much larger in \cabesi\ than in \mgbtwo\ (1.165 and 0.7 st/[eV cell] respectively, see Fig.~\ref{bands}) and the nesting enters in the linewidth expression\cite{allen72} without being renormalized by N(\EF), still we have  {\small $\gamma^{\cabesi}_{E_{2g}}<<\gamma^{MgB_2}_{E_{2g}}$}. 
%%Again, as pointed out by Mazin in the context of the origin of charge density wave \cite{mazinest} the nesting seems to play a not very relevant role here.
%We conclude that the E$_{2g}$   ME {\small $|g^{nn'}_{{\bm k,\bm k'},E_{2g} }|^2$}~ {\it themselves} are very small in \cabesi.  

% and it is confirmed \s\ charge localization in LiBC, where the e-ph coupling is comparable to the one of \mgbtwo. 

The calculated total e-ph is $\lambda$ = 0.38, which makes 
\cabesi\ a weak coupling superconductor, comparable to Al or Mo, 
with a \tc$= 0.4$~K. The two-band resolved values (see Eq. \ref{lph}) 
are $\lambda_{\sigma\sigma}=0.29$; 
$\lambda_{\sigma\pi}=0.21$; $\lambda_{\pi\sigma}=0.15$; 
$\lambda_{\pi\pi}=0.12$ whereas the \mgbtwo\ values are\cite{mgb2PC} 
$\lambda_{\sigma\sigma}=0.83$; 
$\lambda_{\sigma\pi}=0.22$; $\lambda_{\pi\sigma}=0.16$; 
$\lambda_{\pi\pi}=0.28$.  The comparison further shows the 
dramatic reduction of the \s-\s\ coupling in \cabesi.

Although weaker than in \mgbtwo, the contribution from the E$_{2g}$ 
mode is important also in the \cabesi\ 
Eliashberg function (Fig.~\ref{phononeli}), 
where we see a peak at $\sim60$ meV which strongly enhances 
a corresponding structure in the phonon density of states. 
This peak gives roughly one fourth of the total e-ph coupling, 
while the main contribution to the coupling comes from low frequency modes.

We note that  charge localization not only enhances the e-ph interaction,
 but also the Coulomb ME \cite{nota2}, which  are lower in \cabesi\
  than in \mgbtwo. However, this reduction is not so 
  dramatic as for the e-ph ME. This is due to the different structure 
  of the ME of the two interactions (see discussion below).

%In general this prevents even a qualitative prediction of SC properties based only on charge localization arguments, making necessary a full ab-initio calculation.

\section{SUPERCONDUCTING STATE}
\label{superres}

\begin{figure}[t]
  \begin{center}
    \includegraphics[clip,width=0.43\textwidth]{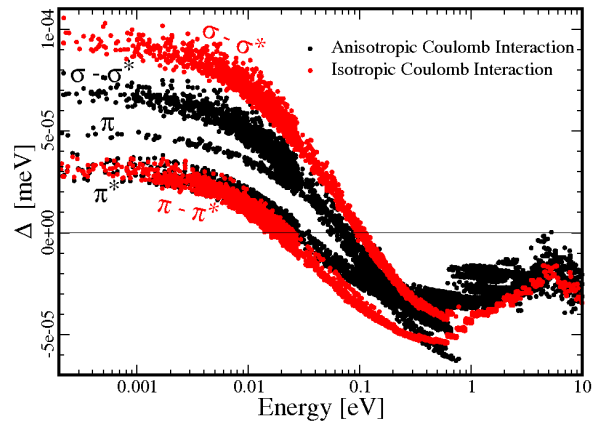}
%    \vspace{0.2cm}

%    \includegraphics[clip,width=0.43\textwidth]{fig2.eps}
  \end{center}

  \caption{(color online) \cabesi\ superconducting gap as a function of the energy distance from the Fermi level.}
  \label{gap}
\end{figure}

The solution of the self-consistent gap equation (Eq.~(\ref{gapeq})) 
including the anisotropic e-ph ME $~\gkkp$ and, 
 the $\nk$-resolved RPA Coulomb matrix elements 
 $V_{n {\bm k},n' {\bm k'}}^{\rm e-e}$, reveals an 
 unexpected complex structure with clearly separated 
 {\it three} gaps at \EF\ (Fig.~\ref{gap}). The calculated critical temperature
 is very low (T$_c$= 0.4), lower than the upper limit (4.2 K) set 
by the experimental results\cite{sano}.
%\begin{multline}%\label{eq:rpame}%{\cal K}_{n {\bm k},n' {\bm k'}}^{\rm e-e}=\sum_{{\bm G},{\bm G}'} \epsilon^{-1}\!\left({\bm q},{\bm G},{\bm G}'\right)\times\\\times4\pi\frac{\left< n'{\bm k}'|e^{i\left({\bm q}+{\bm G}\right)\cdot {\bm r}}|n {\bm k} \right >\left < n{\bm k}|e^{-i\left({\bm q}+{\bm G}'\right)\cdot {\bm r'}}|n'{\bm k}'\right>}{\left|{\bm q}+{\bm G}\right|\left|{\bm q}+{\bm G}'\right|}\,,%\end{multline}
Unlike in \mgbtwo, in which  superconductivity is interpreted  within a 
two-band model\cite{liu,brinkman,golubov,gonnelli_2gap}, 
in \cabesi\ there is a further  \pb-\pa\ gap splitting. 
As in \mgbtwo, the largest gap is related to the \s\ FS sheets 
(cylindrical-like structures in Fig.~\ref{fs}), 
the intermediate one to \pb\ sheets (small hole spheres) 
and the lowest to \pa\ sheets.
The additional \pb-\pa\ gap splitting is a peculiar feature of \cabesi\ not
present in \mgbtwo\ in which the two \s\ and the two \p\ gaps merge together.
In order to understand the origin of this splitting we perform 
some additional computational experiments, solving the gap equation 
$(i)$ completely  neglecting the Coulomb interaction,   
$(ii)$ including only the averaged Coulomb term 
\begin{multline}
\label{a2fcoul}
V_{av}^{\rm e-e}(\epsilon,\epsilon')=\\\frac{1}{N(\epsilon)N(\epsilon')}\sum_{n{\bm k},n'{\bm k'}}V_{n {\bm k},n' {\bm k'}}^{\rm e-e}\times \delta(\epsilon_{\nk}-\epsilon)~ \delta(\epsilon_{\nkp}-\epsilon')\,.
\end{multline} 
and
$(iii)$ with isotropically averaged   Coulomb and 
phononic interactions, corresponding to the dirty limit.

In both $(i)$ and $(ii)$ cases, the three gap structure is destroyed, 
bringing back to a two-band, \mgbtwo-like gap structure. 
In case $(iii)$, instead, superconductivity is completely lost. As a result, 
we predict  
superconductivity in \cabesi\ {\it only} 
if the anisotropic structure of the interactions is included.    
In the real system, very likely, disorder on the Si-Be sublattice can produce
interband  \p-\s\ impurity 
scattering, therefore reducing \tc further.

In the following we will analyse the three gap structure, 
with particular emphasis in understanding the \pb-\pa\ splitting 
observed.  To this purpose, we perform a four-band model analysis, 
splitting the Fermi surface
in: internal \s\ band ($\us$), external \s\ band ($\ds$), \pb\ and \pa\ bands. 

We have calculated (see Table \ref{tab2} $a$ and $b$~) 
the corresponding density of states $N_n$ and the 
BCS-like e-ph couplings, $\lambda_{nn'}$ and $V^{e-ph}_{nn'}$, where: 
\begin{equation}
\label{lph}
\lambda_{nn'}=V^{e-ph}_{nn'}N_{n'}
\end{equation}
 and
\begin{multline}
 \label{vph}
V^{e-ph}_{nn'}= \frac{2}{N_n N_{n'}} \sum_{{\bm k}\in n,{\bm k'}\in n' }
\sum_{\nu} \frac{\gkkp}{\omega_{{\bm k'-\bm k},\nu}} \\
  \times  \delta(\varepsilon_{\nk}-E_F)  \delta(\varepsilon_{\nkp}-E_F).% = 2 <<\sum_{\nu} \frac{\gkkp}{\omega_{{\bm k'-\bm k},\nu}}>>_{FS}
\end{multline}
 First, we notice that the \s\ submatrix $V^{e-ph}_{\sigma\sigma}$ 
 is very homogeneous. Second, the \s-\p\ scattering gives the same 
 contribution to both \s\ gaps, which are then identical.
Superconductivity in the \p\ states is more complicated, essentially
because, unlike \s\ bands,  \pa\ and \pb\  bands originate 
from different sublattices. As a consequence, the \p\ submatrix 
is not homogeneous. Moreover, the \p-\s\ interaction is of the 
same size of the \p-\p\ one, and \pa\ and \pb\  have different 
density of states, being $N_{\pi_a}\simeq 6\cdot N_{\pi_b}$. 
%We now consider a BCS-like model: we start from a ${\bf k}$-resolved BCS gap equation %\cite{BCS} 
%and, using the couplings from Table \ref{tab2}$b$, we impose a four gap solution:  
%$\Delta_{\sigma_1}$,$\Delta_{\sigma_2}$,$\Delta_{\pi_b}$,$\Delta_{\pi_a}$. Moreover, we neglect every anisotropy effect in the quasiparticle energies $E_{n{\bf k}}$\cite{commento}. In this approximation the ${\bf k}$ variable can be integrated out leading to the simplified equation:
%\begin{equation}\label{bcs}
%\Delta_n=C\sum_{n'}\lambda_{n,n'}\Delta_{n'},
%\end{equation}
%where C is a constant coming from the ${\bf k}$ integration.
%Due to the small  $N_{\pi_b}$, all terms in $\lambda_{n,\pi_b}$ can be neglected, and through Eq. \ref{bcs} we see that $\Delta_{\pi_b}$ and $\Delta_{\pi_a}$ are sustained by a different kind of scattering: $\Delta_{\pi_b}$ by \pb-\s\ and the interband \pb-\pa\ ; $\Delta_{\pi_a}$ 
%by \pa-\s\ and the intraband \pa-\pa\ .
%Despite this fundamental difference, the two \p\ gap values, in such a phonon-only model, turn out to be roughly the same, as showed by the SCDFT result performed neglecting the Coulomb terms.
In order to understand the qualitative structure of the SCDFT results, we 
considered a BCS-type approximation of the 4-band Eliashberg
equations\cite{eliash}.
%EVENTUALMENTE QUI SI PUO' METTERE LA FORMULA..
This model calculation confirms that the inclusion of the average Coulomb interaction 
(Eq.~\ref{a2fcoul}), does not produces \pb-\pa\ gap splitting.
In fact, the splitting is recovered  {\it only} considering the 
(band) anisotropic  $V^{e-e}_{nn'}$ Coulomb ME, reported in 
Table \ref{tab2}$c$. 
These are, as for the el-ph coupling, roughly homogeneous in the \s\ 
submatrix and therefore not able to split the \s\ gap, but 
 in the $V^{e-e}_{\pi\pi}$\ submatrix, the intraband interaction 
is $\approx 2.7$ times larger than the interband interaction 
(that couples states in different sublattices).
% and also extremely strong with respect to the e-ph matrix elements (although a direct comparison is not possible because these matrix elements do not include any renormalization effect that is likely to reduce them to about one half attne WHY WE MENTION THE COMPARISON IF IT IS NOT REALLY POSSIBLE TO DO IT?). 
The huge difference between $V^{e-e}_{\pi_b\pi_a}$\ and 
$V^{e-e}_{\pi_a\pi_a}$\, and the low $N_{\pi_b}$ 
(that  makes the $\mu_{\pi_b\pi_b}$ and $\mu_{\pi_a\pi_b}$ negligible), 
lead to a much stronger
suppression of $\Delta_{\pi_a}$ relative to $\Delta_{\pi_b}$, 
eventually bringing to the three gap structure.

%While $\Delta_{\pi_b}$ is not reduced by the strong $V^{e-e}_{\pi_b\pi_b}$ term (because of the low $N_{\pi_b}$ entering $\mu_{\pi_b\pi_b}$), $\Delta_{\pi_a}$  is strongly suppressed (the much larger $N_{\pi_a}$ enters $\mu_{\pi_a\pi_a}$).
%This band anisotropy of the Coulomb interaction, implied by  $N_{\pi_a} > N_{\pi_b}$, is therefore the main cause of the $\Delta_{\pi_b}$-$\Delta_{\pi_a}$ splitting and of the three gap superconductivity predicted for \cabesi. Its origin de%rive, see Fig. \ref{charge}, from the spatial separation of the \pb\ (on Si atoms) from the \pa\ (on Be atoms) charge attne WHAT DOES IT MEAN THIS SENTENCE???. 

\begin{table}
\begin{center}
%{\large ! IF U WANT U CAN TURN ON THIS. THE WIDTH WILL BE STILL OK
\begin{tabular}{ccccc}
\hline
\hline
$a$\hspace{0.3cm}{\bf dos}\hspace{0.5cm} &  $\sigma_1$ & $\sigma_2$ &   $\pi_b$  & $\pi_a$  \\
\hline
$N$        &       0.12       &        0.38      &         0.1      &      0.58           \\ \hline\vspace{0.2cm}
           &                  &                  &                  &                     \\
\hline
\hline
$b$\hspace{0.2cm}{\bf e-ph}\hspace{0.5cm} &  $\sigma_1$ & $\sigma_2$ &   $\pi_b$  & $\pi_a$  \\ 
\hline   
$\sigma_1$ &$\ant{0.07}{0.62}$&$\ant{0.23}{0.61}$&$\ant{0.03}{0.28}$&$\ant{0.20}{0.34}$ \\ \vspace{0.2cm}
$\sigma_2$ &$\ant{0.07}{0.61}$&$\ant{0.22}{0.57}$&$\ant{0.02}{0.23}$&$\ant{0.19}{0.32}$ \\ \vspace{0.2cm}
$\pi_b$      &$\ant{0.03}{0.28}$&$\ant{0.09}{0.23}$&$\ant{0.03}{0.35}$&$\ant{0.16}{0.28}$ \\ \vspace{0.2cm}
$\pi_a$    &$\ant{0.04}{0.34}$&$\ant{0.12}{0.32}$&$\ant{0.03}{0.28}$&$\ant{0.07}{0.12}$ \\ \hline\vspace{0.2cm}
           &                  &                  &                  &                     \\
\hline
\hline
$c$\hspace{0.2cm}{\bf el-el}\hspace{0.5cm}  &  $\sigma_1$  & $\sigma_2$  &   $\pi_b$   & $\pi_a$  \\ 
\hline   
$\sigma_1$ &$\ant{0.11}{0.91}$&$\ant{0.27}{0.71}$&$\ant{0.03}{0.30}$&$\ant{0.15}{0.26}$ \\ \vspace{0.2cm}
$\sigma_2$ &$\ant{0.08}{0.71}$&$\ant{0.32}{0.84}$&$\ant{0.03}{0.29}$&$\ant{0.13}{0.23}$ \\ \vspace{0.2cm}
$\pi_b$      &$\ant{0.04}{0.30}$&$\ant{0.11}{0.29}$&$\ant{0.09}{0.94}$&$\ant{0.20}{0.35}$ \\ \vspace{0.2cm}
$\pi_a$    &$\ant{0.03}{0.26}$&$\ant{0.09}{0.23}$&$\ant{0.03}{0.35}$&$\ant{0.51}{0.86}$ \\ 
\hline 
\end{tabular}
%}

\caption{Table $a$: Band resolved density of states [st/(cell$\cdot$eV)]. Table $b$: Band resolved e-ph couplings $\lambda_{nn'}$ (in bold) and BCS effective potentials $V^{e-ph}_{nn'}=\lambda_{nn'}/N_n'$ (in parenthesis [eV]). Table $c$: Band resolved el-el interactions at the Fermi energy $\mu_{nn'}$ (in bold), and  el-el matrix elements $V^{e-e}_{nn'}$ at the Fermi energy (in parenthesis [eV]).\label{tab2}}
\end{center}
\end{table}

\section{SUMMARY AND CONCLUSIONS}
\label{concl}

We have calculated the normal and superconducting state
properties of \cabesix\ ($x=1$) in the {\it \albtwo} phase,
 within the density functional theory for superconductors  (SCDFT). 
 The chosen doping level is not far from the experimental doping 
 ($x=0.75$) where this phase is found stable 
and homogeneous. \cabesi\ is isostructural and isoelectronic to 
\mgbtwo\ and the electronic, vibrational properties and electron-phonon 
 interaction of the two materials
are compared directly: While the band structures present 
strong similarities, with both \s\ and \p\ bands crossing 
the Fermi level, the phonon structure and the e-ph interaction
 differ substantially. In particular,  the less localized \s\ 
 charge of \cabesi\ brings to a dramatic reduction in the $E_{2g}$ 
 electron-phonon coupling, with a consequent reduction of the phonon 
 renormalization seen in \mgbtwo. This fact makes \cabesi\ a weak
coupling superconductor with e-ph $\lambda= 0.38$ and \tc$=0.4$~K,
in spite of a nesting at the Fermi surface twice as big as in \mgbtwo\ .
Interestingly, \cabesi\ exhibits three superconducting gaps at the
 Fermi level. While, as in \mgbtwo, the \s-\p\ gap 
 splitting is related to the different e-ph coupling in these bands, 
 the further \pb-\pa\ splitting  is a pure effect of the complex 
 structure of the anisotropic Coulomb repulsion, acting in different 
 way on the  \pb\ and \pa\ states.

%\cabesi\ can be view as a MgB2 with the relevant 
%parameters changed in a realistic way. In fact, 
%the analysis of the Fermi surface, the phonon structure and the e-ph coupling,
% as compared with the one of \mgbtwo, reveals once again the subtle nature of 
%the SC state, with the dramatic change of \tc even when atomic 
%substitution (leaving the structure and formal valence unchanged)
%are made. 

\section{ACKNOWLEDGMENTS}
\label{ackno}

The Authors thank A. Marini for technical support and for making the SELF code available. 
This work was  supported by the Italian Ministry of Education, through a 2004 PRIN project, by the Istituto Nazionale di Fisica della Materia (INFM-CNR) through a supercomputing grant at 
Cineca (Bologna, Italy),  by MIUR under project PONCyberSar, and by the Deutsche Forschungsgemeinschaft, by the EXCITING Network and by the NANOQUANTA Network of Excellence. 
A.F. acknowledges CNISM for financial support during his visit in University of L'Aquila. 
A.S. acknowledges  the \textit{Regione Autonoma della Sardegna} for providing the Master and Back fellowship. C.B. acknowledges for the financial support of the Swiss National Science Foundation.


\begin{thebibliography}{99}
\bibitem{nota} The chosen doping $x=1$  is not far from the experimental stoichiometry ($x=0.75$) and  avoids  cumbersome supercell approaches and problems related to Be and Si ordering. %,  not affecting the conclusions of the work.
\bibitem{nota2} In \mgbtwo\ for example, the strong hole-doped covalent B-B bond  ensures an high el-ph coupling: however, an equally high repulsive Coulomb interaction reduces the superconducting critical temperature.
\end{thebibliography}
\end{document}